\documentclass[conference]{IEEEtran}
\usepackage[utf8]{inputenc}
\usepackage{graphicx}
\usepackage{subfigure}
\usepackage{fancyhdr}
\usepackage{xcolor}
\date{January 2020}

\begin{document}

\title{Trust-by-Design: Evaluating Issues and Perceptions within Clinical Passporting}

\author{\IEEEauthorblockN{Will Abramson\IEEEauthorrefmark{1}, Dr Nicole E. van Deursen\IEEEauthorrefmark{2}, William J Buchanan\IEEEauthorrefmark{1}}
\IEEEauthorblockA{\IEEEauthorrefmark{1}Blockpass ID Lab, School of Computing, Edinburgh Napier University, Edinburgh, UK.
\\\{1\}@w.abramson,w.buchanan@napier.ac.uk}
\IEEEauthorblockA{\IEEEauthorrefmark{2}National Cyber Security Centre, The Hague, The Netherlands, Edinburgh Napier University, Edinburgh, UK.
\\\{2\}n.vandeursen@napier.ac.uk}
}

\maketitle

\begin{abstract}
A substantial administrative burden is placed on healthcare professionals as they manage and progress through their careers. Identity verification, pre-employment screening and appraisals: the bureaucracy associated with each of these processes takes precious time out of a healthcare professional's day. Time that could have been spent focused on patient care. In the midst of the COVID-19 crisis, it is more important than ever to optimize these professionals' time. This paper presents the synthesis of a design workshop held at the Royal College of Physicians of Edinburgh (RCPE) and subsequent interviews with healthcare professionals. The main research question posed is whether these processes can be re-imagined using digital technologies, specifically Self-Sovereign Identity? A key contribution in the paper is the development of a set of user-led requirements and design principles for identity systems used within healthcare. These are then contrasted with design principles found in the literature. The results of this study confirm the need and potential of professionalising identity and credential management throughout a healthcare professional's career.

\end{abstract}

{\bf Keywords:}
COVID-19, CL-15, CL-17, privacy-preserving,

\section{Introduction}

While the COVID-19 crisis has brought the challenges of staff mobility into the spotlight, the administrative burden placed on a healthcare professional throughout their career has always been present. Over the years healthcare service providers have increased the minimum standard for identity verification and pre-employment checks in line with new regulations \cite{regulatory_overload, ruscoe_identity_2018}. As a result, the time spent on these processes has increased. A House of Lord's report, for example, estimated that 25,000 junior doctor days a year were currently being spent on these administrative tasks \cite{holmes2018distributed}. In addition to this, digitisation of healthcare services has further increased the time and complexity associated with managing one's career. In a 2011 US survey, 87\% of physicians stated that the leading cause of stress was down to administration \cite{keswani2011increased} and a study of Finnish physicians found that poorly functioning IT systems continue to be a major cause of stress, particularly for those in highly time pressured roles \cite{heponiemi_finnish_2017}. Another major challenge is within doctors working across both in public health care and also in private practice \cite{peisah2009secrets}.

In a crisis like the COVID-19 outbreak, the need for a healthcare service to react to rapidly evolving, location specific, stresses at a Trust or Hospital level cannot be clearer. Different locations may hit their peak at different times, while some areas may only be minimally affected \cite{ferguson2020report}. However, consultation with an RCPE trainee suggests that on-boarding into a new Trust or hospital can take up to two days. This is two days of precious time that could potentially have been spent saving lives. Technological solutions have regularly been heralded for their ability to reduce inefficiencies and streamline patient care. Blockchain technology is just one of the more recent innovations predicted to have a disruptive impact \cite{mettler2016blockchain}. Often though, the reality in the hospitals is different to the design assumptions made by technologists and the productivity benefits are not always obvious \cite{thouin2008effect}. 

This paper thus presents an initial set of design principles for any technical solution attempting to reduce the administrative burdens currently placed on healthcare professionals. An analysis of discourse about digital identities, verifiable claims and trust has led to a theoretical set of trust and design principles. These principles were validated in a workshop with healthcare organisations held at the RCPE. This research takes an initial step towards understanding the problem space from the perspective of those currently experiencing it and lays the foundation for future quantitative studies in this area.

\subsection{Research questions}
This paper evaluates a use case in which a person can digitally obtain, manage and present their professionals credentials and personally identifiable data within an healthcare system. We limited the scope of the work to healthcare professionals, as these are identified as being burdened with administrative tasks associated with identity verification, and pre-employment checks. As well as recording and managing their credentials as they progress through their career. A burden which is generally expected to take place in people's personal lives. The following research questions were identified:

\begin{itemize}
    \item What are the identity interactions that a healthcare professional must manage throughout their career?
    \item How might Self-Sovereign Identity technology be used to simplify a healthcare professionals identity administration?
    \item How do the design principles of Self-Sovereign identity stated in the literature and technical sphere meet the requirements of the healthcare professionals that would actually be using these systems?
\end{itemize}{}




\section{Related Literature}
Berwick, Nolan and Whittington define the \emph{Triple Aim} focusing on improving the care, health and cost when accessing healthcare performance in the US \cite{berwick2008triple}. They point out that these goals are interdependent so must be considered together when planning and evaluating healthcare changes. It has been suggested that this framework should be extended to consider a fourth aim, care for the provider \cite{bodenheimer2014triple}, due to reports that staff burnout and dissatisfaction is widespread. As care providers are on the frontline when it comes to achieving the triple aim for healthcare services, including their well being into this assessment makes sense.





\subsection{Healthcare Professional Credentials}

Healthcare providers have a requirement to maintain strong identity verification checks to ensure that their employees are who they claim to be and that they have the required skills and training for the job \cite{ruscoe_identity_2018, fsmb_report, nhsemployers}. Unfortunately, there have been examples throughout the world of doctors practising without licences. This puts patients lives at risk and reduces the trust in the profession as a whole. Examples include; the UK General Medical Council recently having to recheck credentials of 3,000 doctors after a fraudulent psychiatrist was found to have practiced for 23 years without proper credentials \cite{dyer2018gmc}, the case of a social worker in Canada involved in more that 100 child protection cases \cite{cred_fraud_canada} and the notorious US case of Christopher Duntsch a.k.a Dr Death \cite{dr_death}.

As a consequence, credentialing healthcare professionals is a crucial process in healthcare systems throughout the world. However, the current practices of many systems add huge overheads to both the administrators and the healthcare professionals. In a report on healthcare and digital credentials the US Federation of State Medical Boards (FSMB) \cite{fsmb_report} analyse the use of digital credentials in healthcare, looking at the potential for both current technology and future technology to streamline the process and enhance trust in the system. The implementation of the Federation Credentials Verification Service (FCVS) in 1996 \cite{fsvc}, a NCQA-certified platform providing a centralized service for obtaining primary source, verified education information for medical practitioners applying for licensing in the US. As the report \cite{fsmb_report} outlines, the FCVS reduced the time to obtain a license from 60 days to ~25 days. A significant reduction. However, efforts to improve the FCVS highlighted the underutilization of technology in the process. Furthermore 66\% of this time is driven by parties outside of the control of the FCVS \cite{fsmb_report}. These credential verification organisations are often redundant and increase the cost of the whole process. The report highlights the movement to \textit{disintermediate the creation and management of credentials}, hinting at a movement towards individual ownership of credentials.

 Along with this there is an increasing need for clinical staff to provide digital evidence of their training, skills and experience. Read et al \cite{read2017clinical} investigated the usage of a passporting system for surgery clerkship and found that those involved often found that it improved student's reporting of their performance in basic clinical skills.


\subsection{Self-sovereign identity (SSI)}

Digital identifiers - and the trust entities place in them - enable many modern societal interactions. They thus allow organisations to perform critical activities with increased levels of trust. Another way of looking at it is from a risk perspective, and where digital identifiers and account information correlated with an identifier that helps organisations make risk-based decisions associated with a particular interaction and value exchange. Unfortunately, traditional identity management systems continue to have security risks \cite{cser_forresters_2017}, such as: credential theft or loss; biometric impersonation; document forgery; and identity theft. 

SSI uses a new type of identifier currently going through standardisation at the W3C, a Decentralised Identifier (DID) \cite{dids}. A DID is an identifier under the sole domain an entity, typically the entity that created it. They are cryptographically verifiable and independent of any central authority. Rather than being assigned an identifier on account creation, DIDs let individuals provide their own identifiers for their digital relationships.For a more detailed description of the technical architecture pf self-sovereign identity we point the reader to the following survey \cite{muhle2018survey}.

Systems built following an SSI architecture could offer the opportunity to rethink the entire credential process for physicians. Before the electronic transmission of credentials can be put forward as a viable option for healthcare professionals, it must be considered if such a process would break any of the rules and regulations currently governing this area. The key things identity verification and authentication process must satisfy for most healthcare services are \cite{fsmb_report, ruscoe_identity_2018}:
\begin{itemize}
    \item Is it possible to verify the authenticity of the claim?
    \item Is the claim a primary source attestation. For example, is your degree certification a certificate from the university you attended?
    \item Was the credential securely delivered from the credential holder to the verifier?
    \item Is there a clear, verifiable audit trail that can trace the origin of this credential?
\end{itemize}

These points can, in fact, all be satisfied digitally through the use of digital signatures.

The problem of scaling digital identities into health care systems have lead many people to explore alternative methods to identify and authenticate people and things in the digital sphere. One of the proposed architectures is commonly referred to as Self-Sovereign Identity (SSI). Connor-Green identifies that SSIs could be one of the core use cases of blockchain in health \cite{connor2016blockchain}. Liang et al outline a blockchain approach within a heath care management system and where the distributed nature of SSIs support a scalable infrastructure which moves away from the centralised control of identity within many existing health care infrastructures \cite{liang2017towards, liang2017integrating}.

\begin{figure*}
    \centering
    \includegraphics[width=0.8\linewidth]{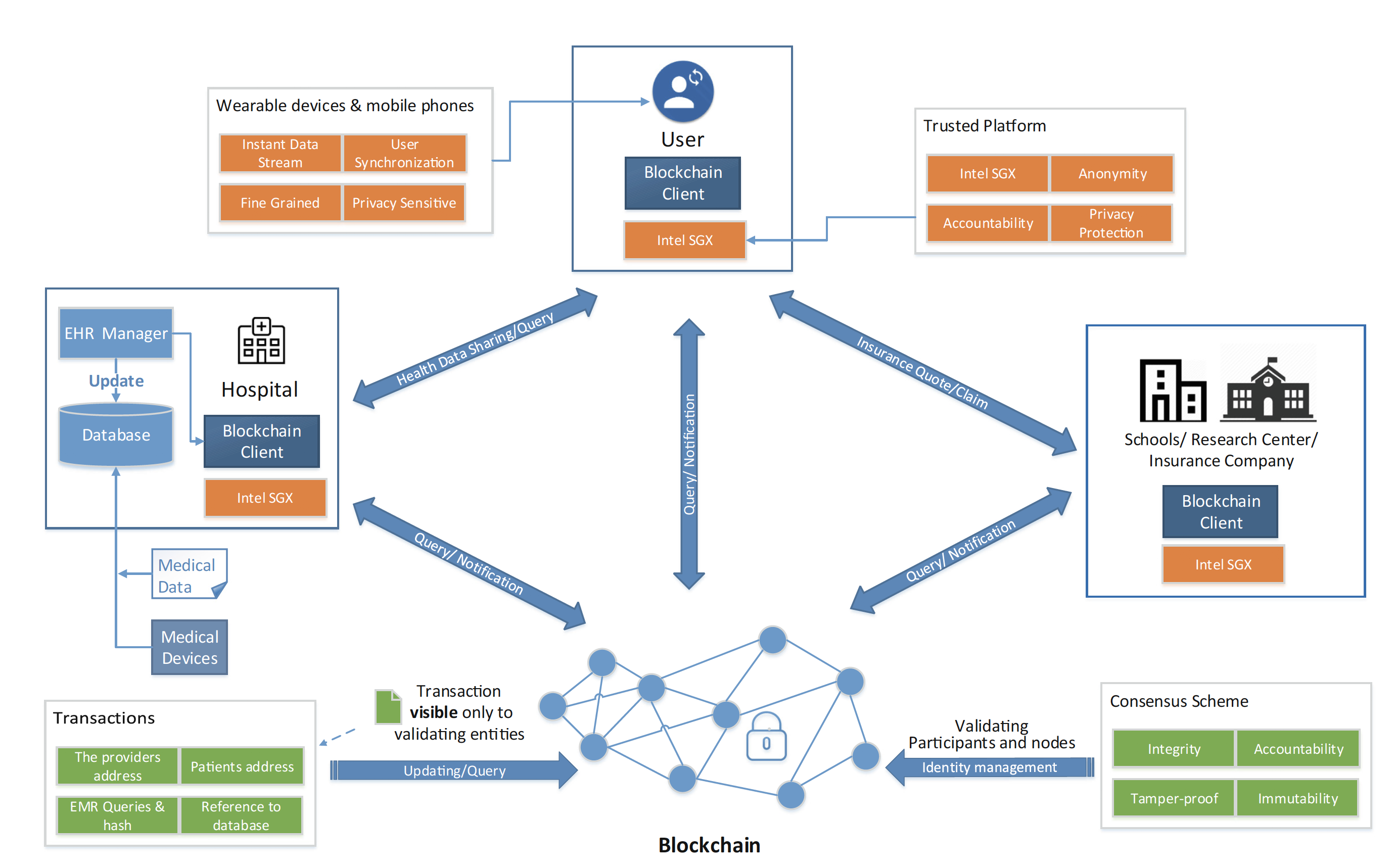}
    \caption{Patient centric personal health data management system \cite{liang2017towards}}
    \label{fig:centric}
\end{figure*}


\section{Method}


The goal of the workshop was to gather a set of values and principles that can be used to evaluate emerging technology for identity systems within healthcare. Terminology and definitions are often much contested in different contexts. We used definitions provided in the selected papers as much as possible, but some definitions were adapted to better fit our goal to measure the value or principle.

The next step in our research was a workshop organised at RCPE. It involved 14 participants with a wide ranging experience of different aspects of the healthcare system. The participants for the workshop were selected through consultation with the RCPE, and who were able to use their contacts to invite a diverse range of attendees. This included clinicians, RCPE trainees, and RCPE staff involved in data management, digital transformation and education, as well as a representative from the General Medical Council (GMC). While no personal data was captured during the workshop and all attendees remain anonymous, explicit consent was obtained and the research aims were explained at the beginning of the workshop.

During the workshop, the participants evaluated the fit of SSI within the healthcare domain. First, a process mapping exercise was used to develop an understanding of the current system, looking at the identity data exchanges and key entities that verify or attest to identity attributes of healthcare professionals throughout their career.

Then these identity moments were re-imagined within a SSI enabled ecosystem, this story was told in an interactive manner using physical props and audience participation in order to convey the capabilities that SSI systems could provide for healthcare professionals, without going into unnecessary technical detail. Finally, the workshop participants were asked to evaluate the positive and negative aspects of the identity management system. The participants expressed their requirements, values and expectations. The list that was distilled from these workshop discussions was compared against the list of design principles from the desk research.

\section{Design principles for self-sovereign identity systems}

The success of any SSI system depends not only on the technical feasibility but also on the user acceptance and trust in the system. With users we mean all stakeholders (entities) within a specific context that will use that system together. Trust is harder to define, as there exist more than 70 definitions in academic literature \cite{seppanen2007measuring}. Trust is seen as a human strategy to cope with uncertainty, such as those we face in relations, actions and innovation \cite{van2017regulating}. In a digital context, trust is often transferred to cybersecurity measures such as technological controls, certificates, and organisational compliance frameworks. In SSI systems, at least some aspects of this trust shifts from trust between people towards confidence that is placed in cryptographic systems. As Smolenski \cite{smolenski_evolution_2018} frames it: \emph{trust is being depersonalized}. Designing new digital identity systems means mimicking real-life situations in a digital way. Human values, such as ethics and trust need a digital equivalent that users need to accept and understand. How do we design trust from the start? Which design principles are most valued by users and most likely to establish trust in these systems?


There are many papers that refer to the \emph{Laws of Identity} (coined by Cameron in 2005) \cite{cameron_laws_2005} as the foundation for design principles. These laws explain the dynamics causing digital identity systems to succeed or fail in various contexts. Although written before the era of SSI, Cameron himself finds the laws still relevant, also for identity systems on the blockchain and decentralized identity \cite{cameron_laws_2018}. He points out, for instance, that the first four of the laws are also requirements within the GDPR. 

In 2016, Christopher Allen \cite{allen_path_2016} wrote 10 principles inspired by (amongst others) the work of Cameron. His aim was to ensure that user control is at the heart of SSI. Allen pointed out that identity can be a double-edged sword: it can be used for both beneficial and maleficent purposes. Therefore, he states: \emph{an identity system must balance transparency, fairness, and support of the commons with protection for the individual.} The Sovrin Foundation \cite{tobin2016inevitable} adopted Allen’s principles and arranged them in three sections, but this causes some confusion as they used one principle twice and made another principle a section above other principles. Other researchers and developers have used Cameron’s or Allen’s principles for inspiration and adapted them to their own lists of features. However, testing of SSI systems against these features and design principles is still rare. 

Dunphy \& Petitcolas \cite{dunphy2018first} evaluated three identity management solutions (uPort, Sovrin, and ShoCard) against Cameron’s laws of identity. Their overview shows that none of the solutions meets all seven laws, and non of them meets the law of human integration: usability, user understanding and user experience. They state that none of the schemes they evaluated are accompanied by an evidence-based vision of user interaction. One of the limitations is usable end user key management for nontechnical users that remains unaddressed. Furthermore, they express concern about tightening regulation, such as the GDPR, that sometimes contradicts the transparency of data storage in these solutions. Finally, most solutions provide only ad-hoc trust, as trust relies on integration between participating entities and methods to achieve trust in the context of identity attributes are still evolving. 

Ferdous et al. \cite{ferdous2019search} elaborated on the principles of Allen and designed a taxonomy of essential properties for SSI. Then, they compared four blockchain-based SSI systems (uPort, Jolo, Sovrin, Blockcerts) against the properties and through desk research they found that most of the systems satisfy most of the properties. Similar work was done in a student project \cite{van2019self} where students compared eight blockchain-based (IDchainz, Uport, EverID, Sovrin, LifeID, Selfkey, Shocard, Sora) and three non-blockchain based SSI systems (PDS, IRMA, reclaimID) against each of Allen’s principles with one additional principle \cite{stokkink2018deployment}. They concluded that some of the blockchain based solutions fulfil all properties, but that some of the non-blockchain-based implementations meet most of the criteria as well. Interestingly, their conclusions as to whether the properties are met do not always match the conclusions of \cite{ferdous2019search} for two systems (Uport and Sovrin) that both projects evaluated. Toth and Anderson-Priddy \cite{toth2019self} validated nine properties from earlier sources (e.g. Allen and Sovrin Foundation) and added new properties. They applied these properties to their architecture for digital identity and reasoned how these apply to their solution (NexGenID).  

To the best of our knowledge, published evaluation of values and principles with users in a specific SSI context is very rare. One project that focused on citizens and digital identity systems in general (not SSI specific) was the Digital Identity lab in The Netherlands. In several interactive sessions with citizens they found which values matter the most for digital identities \cite{spierings_digitale_2019}. The research methods included interviews in the streets, meet-ups, expert sessions and design sprints. The results include evaluation quadrants to plot digital identity providers and an overview of values that citizens find important, and that can be used as input for ethical design and trust of digital identity systems. Another project focusing on user experience is the IRMA Made Easy project \cite{schraffenberger_irma_2020}. IRMA (I Reveal My Attributes) is a self-sovereign identity solution with a digital wallet. The IRMA Made Easy Project works on the design of the app and website with a focus on accessibility. The developers of IRMA point out that user experience design affects how users handle the control over their information. From their experience they share three lessons: 

\begin{enumerate}
\item In order for new technology to be adopted, they require a smooth user experience. 
\item User experience design for privacy is not the same as general user experience design. 
\item A system that puts people in control over their data does not always lead to people using that control to protect their privacy: it can even lead to the opposite when they are tricked by others. 
\end{enumerate}
  
From the literature study we learned that there is a gap in academic research that includes evaluation of proposed SSI solutions from a user perspective with domain knowledge of the ecosystem. Furthermore, to the best of our knowledge, the most commonly used design principles have not been validated by users for importance and priority. Projects that included consumers focus on identity management in general, and studies on SSI systems tend to focus on the evaluation by technical experts through desk research. Furthermore, when SSI design principles and features indeed are evaluated, the researchers re-use existing frameworks or lists of principles without user elicitation for principles that technology experts have not imagined yet. If we do not understand the requirements of end users then we run the risk of creating digital tools that no one wants to use, or worse introducing unintended consequences through the deployment of these systems to domains with poorly understood requirements. There are countless examples of technology being introduced into healthcare only to make the jobs of those working alongside this technology worse, like the 15 logins needed to access different NHS systems \cite{bbc_logins}.

We compared the different lists of principles, features and values that we found, and created an overview of different and overlapping principles. The results are presented in table \ref{tab:principles}. The overview of principles was input for the next stage in our research, where we invited future users to express their opinion on principles and values. In the next section we describe the workshop that we held with representatives of different entities in an ecosystem, in order to contribute to the knowledge of end-user perception and trust of SSI systems.

\begin{table*}[ht]
\label{tab:principles}
\begin{center}
 \begin{tabular}{|p{3cm}|p{3cm}|p{3cm}|p{3cm}|p{3.5cm}|} 
 \hline 
 Cameron  \cite{cameron_laws_2005} & Allen \cite{allen_path_2016} & Ferdous et al \cite{ferdous2019search} & De Waag \cite{spierings_digitale_2019} & Toth Anderson Priddy \cite{toth2019self} \\ [0.1ex] 
 \hline\hline
  & Existence & Existence &  &  \\ 
 \hline
 User Control and Consent & Control and Consent &  Consent &  Control & Control and consent \\
 \hline
  & Access & Access & Access & Access  \\
 \hline
 & Transparency & Transparency & Transparency &  \\
 \hline
Pluralism of operators and technologies & Portability & Portability &  & portability \\
 \hline
 Consistent experience across contexts & Interoperability & Interoperability & & Interoperability\\
 \hline
  & Persistence & Persistence & & Persistence \\
\hline
Minimal disclosure for a constrained use & Minimalization & Minimalization & Data-minimalization & \\
\hline 
 & Protection & Protection & Security  & secure transactions and identity transfer\\ 
\hline
 & & Autonomy & Autonomy & \\
\hline
 Justifiable parties & & Choosability & & \\
\hline
 Human integration & & & Ease of use & usability \\
 \hline
 & & Disclosure &  &  \\
\hline
 & & Ownership & & \\
 \hline
  Directed identity & & Single source & & \\
 \hline
& & Standard & & \\
\hline
& & Cost & &\\
\hline
& & Availability & & \\
\hline
& & & Trust & \\
\hline
& & & Privacy & \\
\hline
& & & Integrity & \\
\hline
& & & Decentralization & \\
\hline 
& & & Inclusivity & \\
\hline
& & & Reliability & \\
\hline 
&  & &  & Counterfeit prevention\\
\hline
& & &  & Identity verification\\
\hline
& & &  & Disclosure\\\hline
& & & & Identity assurance\\[1ex]
 \hline
\end{tabular}
\end{center}
\caption{List of design principles}
\label{tab:principles}
\end{table*}

\begin{table*}
\begin{center}
 \begin{tabular}{|p{3cm}|p{10cm}|} 
 \hline
 Design Principle & Definition \\ [0.5ex] 
 \hline\hline
Existence &	The identity of a person exists independent of identity administrators or providers.
Control	The person is in control of their digital identity and is able to choose what personal data to share.  \\ 
 \hline
Autonomy &	A user is independent on creating identities, as many as required, without relying on any party and be able to update/remove it. \\
 \hline
Disclosure	& A user must have the ability to selectively disclose particular attributes.  \\
 \hline
Ownership &	The user is the ultimate owner of an identity, including the claims. \\
 \hline
Consent	& Data must be released only after the user has consented to do so. \\ 
\hline
Access &	The person has full access to their own data.\\
\hline
Single source &	A user is the single source of truth regarding the identity. \\
\hline
Transparency &	Systems and algorithms are transparent and anyone should be able to examine how they work. \\
\hline
Standard &	An identity must be based on open standards.\\
\hline
Cost &	Costs must be kept to minimum.\\
\hline
Portability &	Information and services about identity must be transportable to other services.\\
\hline
Interoperability &	Digital identities are continually available and as widely usable as possible.\\
\hline
Persistence	& An identity must be persistent as long as required by its owner.\\
\hline
Minimalization	& When data is disclosed, that disclosure should involve the minimum amount of data necessary to accomplish the task at hand. For example, if only a minimum age is called for, then the exact age should not be disclosed, and if only an age is requested, then the more precise date of birth should not be disclosed.\\
\hline
Protection	& The rights of users must be protected. When there is a conflict between the needs of the identity network and the rights of individual users, then the network should err on the side of preserving the freedoms and rights of the individuals over the needs of the network.\\
\hline
Availability & 	An identity must be available and accessible from different platforms when required by its owner. \\
\hline
Human welfare &	The identity system must contribute to human well-being.\\
\hline
Non-maleficence &	The system will not cause harm to others.\\
\hline
Justice	& Systematic unfairness (false negatives/positives) is avoided.\\
\hline
Trustworthiness &	Expectations to act with good will towards others.\\
\hline
Privacy	& The user has the right to decide what data is shared and to set boundaries.\\
\hline
Dignity	& Dignity is intertwined with emotional identity. Technological solutions have a responsibility to uphold human dignity.\\
\hline
Solidarity &	Respectful cooperation between stakeholders.\\
\hline
Environmental welfare &	The solution should cause no environmental harm.\\[1ex] 
 \hline
\end{tabular}
\end{center}{}
\caption{The list of design principles was validated in a workshop}
\label{tab:design_defs}
\end{table*}

\section{Workshop description}

After a brief introduction, participants were asked to complete a warm up exercise where they recorded different identity interactions that occur throughout a typical day in their life. This included using a RFID card, logging into a digital system with a username and password, authenticating to a mobile device, bank card payments. Anything that involved some form of identification and authentication. Participants were also asked to record times when authentication failed for example through a rushed or forgotten password attempt.

The aim of the exercises was to get attendees thinking about how often they interact with digital systems, how many different username and passwords they currently manage and the number of different authentication devices that they have to carry. The majority of participants recorded over 25 separate identity interactions, all in a single day.

\subsection{Healthcare Ecosystem Process Mapping}

For the next stage of the workshop focused on eight core identity moments that captured at a high level the typical experiences of a doctor throughout their career. Participants were asked a create process maps identifying key organisations involved in each of these stages and the identity information that a doctor is required to present to them. Additionally participants were asked to capture frustrations that a doctor might experience while navigating these identity moments. The general identity moments identified and validated prior to the workshop to provide some structure were as follows:

\begin{itemize}
    \item Doctor graduating from university.
    \item Doctor applying for a job.
    \item Doctor joining a hospital.
    \item Doctor training.
    \item Doctor rotation.
    \item Doctor begins RCPE accreditation.
    \item Doctor qualifies as a Physician.
    \item Doctor moves abroad.
\end{itemize}{}

\begin{figure*}
    \centering
    \includegraphics[width=1.0\linewidth]{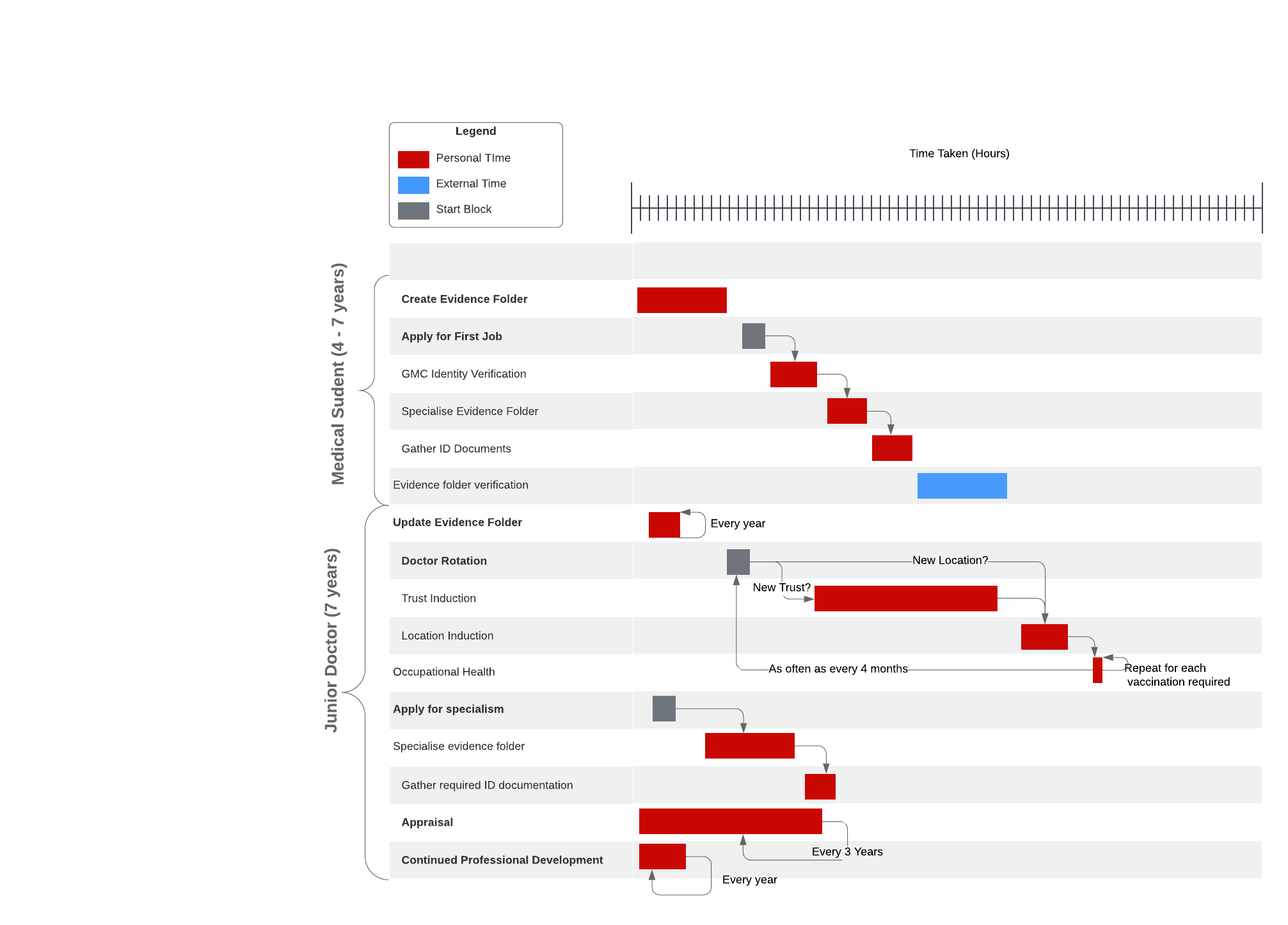}
    \caption{Healthcare Professional's identity moments}
    \label{fig:id_moments}
\end{figure*}

The workshop participants were split into groups, each group focused on four of the identity moments. The results were then presented back to the group providing a detailed overview of each of the stages in a doctor's career, including recurring and trusted ecosystem entities such as the GMC (General Medical Council). These maps were combined and synthesised into a Gantt chart, showing the time burden and repetition associated with a healthcare professionals as they progress through their career, see  Figure \ref{fig:id_moments}. This was developed through follow up communication with a final year trainee at the RCPE who attended the workshop.

\subsection{Re-imagining Identity Moments using SSI}

The next stage of the workshop involved an interactive session on Self-Sovereign identity and the capabilities it could provide healthcare professionals when applied to the key identity moments that participants previously mapped out. The goal was to give attendees a high level understanding of how this technology works and where it might fit into existing processes within healthcare.

Physical props were used to represent different aspects of the SSI system and a number of workshop attendees were asked to play roles within the healthcare ecosystem. Specifically, six participants acted as the key entities and trust providers identified in the process mapping stages: 
\begin{itemize}
    \item \textbf{A Medical School} - Before becoming a licensed doctor, individuals must first complete a degree at a medical school.
    \item \textbf{The General Medical Council} - The doctor licensing body in the UK
    \item \textbf{The Royal College of Edinburgh} - Royal colleges are involved with training and examination procedures for junior doctors as they gradually specialise in a medical discipline.
    \item \textbf{Edinburgh Hospital} - This hospital was used as the initial place of employment once the fictional doctor in our scenario graduated.
    \item \textbf{Glasgow Hospital} - This entity represented the doctor rotation process within the scenario modelled.
    \item \textbf{Health Education Scotland} - A body involved with continuous training and education of doctors.
\end{itemize}{}

The initial setup of the SSI healthcare ecosystem was represented by asking each actor in the scenario to \textit{generate a pubic/private key pair}. A red (private) and green (public) card was used to show the two halves of a public/private key pair. Actors then attached their \textit{public key} to a white card, which was used to represent a decentralised identifier (DID). All actors were asked to place their DID onto the wall, representing the act of registering a public DID on a distributed ledger such that the public keys for these trusted entities could be resolved by anyone.

\begin{figure*}
    \centering
    \includegraphics[width=1.0\linewidth]{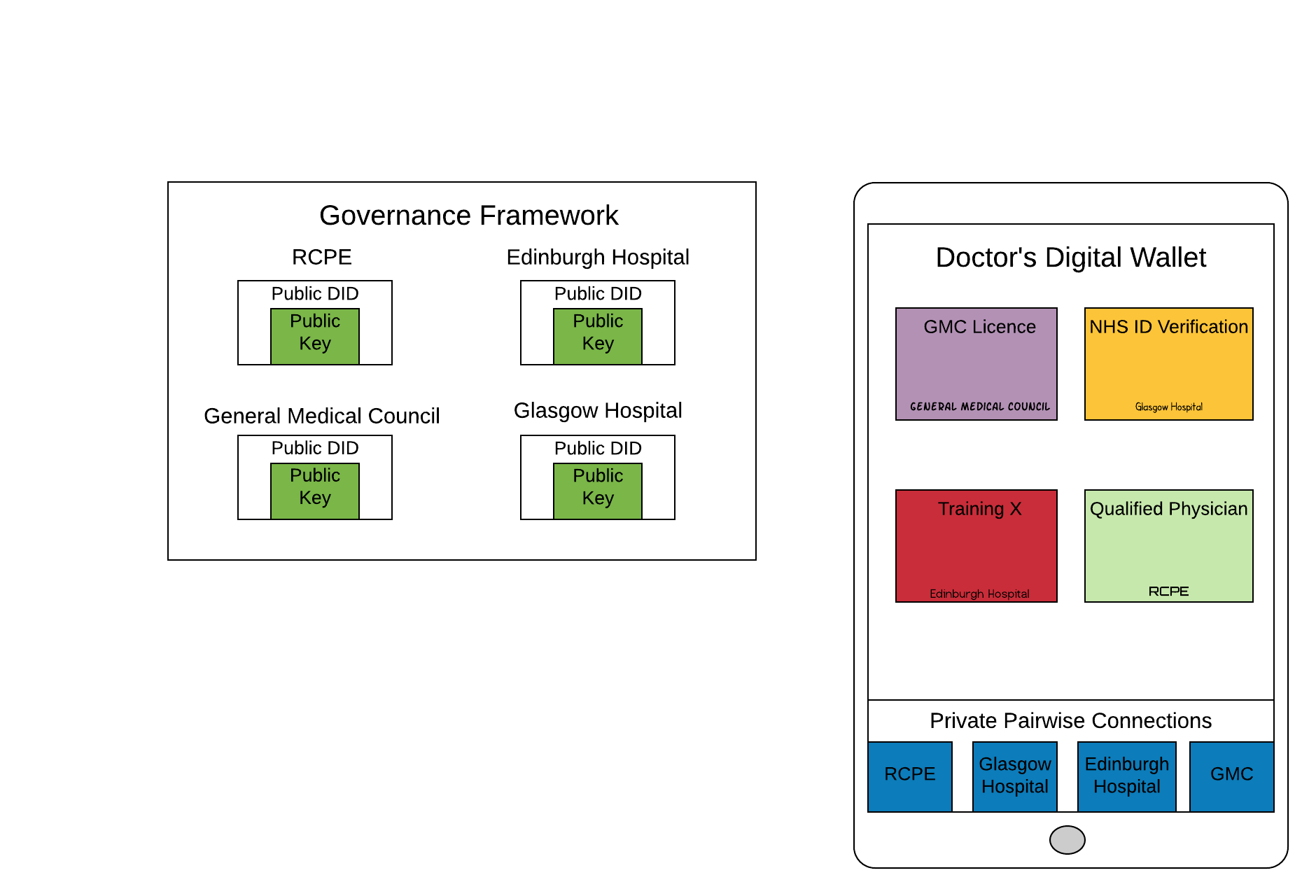}
    \caption{A illustration of the workshop's physical representation of an SSI ecosystem}
    \label{fig:finalsetup}
\end{figure*}{}

After the initial setup, a member of the research team played the role of doctor and walked through each of the identity moments discussed and mapped in the process mapping session. This interactive approach was used for a couple of purposes:
\begin{itemize}
    \item To educate workshop attendees about the capabilities that a Self-Sovereign Identity (SSI) architecture enables.
    \item To illustrate how SSI could be applied to the healthcare domain to streamline identity interactions.
\end{itemize}{}

Throughout this interactive session, a flipchart was used to represent the doctors digital wallet. The wallet gradually collected Verifiable Credentials represented as large post-it notes and digital relationships, formed through peer DID connections, were represented as small blue post-its within the wallet. See Figure \ref{fig:finalsetup}.

Within a SSI system there are generally four core interactions; 

\begin{itemize}
    \item Establishing a peer DID connection to initiate a digital relationship.
    \item Credential issuance.
    \item Proof request.
    \item Presentation of credential attributes in response to a proof request.
\end{itemize}{}

Each of these interactions were represented in this scenario as follows:

\subsubsection{Establishing a peer DID connection}
For this interaction, the action used was simply a handshake between the doctor and the party the doctor wished to connect with. For example, during the doctor qualifies from medical school process mapping the doctor had to establish a connection with the GMC. Whilst the handshake took place participants were explained that this represented forming a private peer-to-peer communication channel across which message integrity and authenticated origin can be verified. It was illustrated that the connection was stored and managed by the digital wallet using a small blue post-it note. 

The fact that these connections could be formed either face to face, or through a website was additionally discussed. Explaining that you probably have more trust in a connection formed face to face and that trust can be built across these connections by sharing verifiable information.

This relationship once formed can last indefinitely, until one individual or the other decides to break it off. This means that the GMC or any entity could form more personal relationships with the doctors they licence, enabling them to push them relevant communication across their secure communication channel.

\subsubsection{Credential Issuance}

Credential issuance was a recurring SSI interaction that was represented in the scenario. The medical school issued the doctor their medical degree, the GMC issued their licence and the RCPE issued the doctors a qualified physician credential. All these credentials exist in the current system. They are attestations about the attributes and qualifications of a doctor, made by entities with the authority to make such claims. Before any credential issuance interaction, a secure connection must have been formed as discussed above this was highlighted through a handshake action.

Then workshop participants acting as the different trusted entities within the scenario were asked the write the name of the credential (a large post-it) and sign the credential using their \textit{private key} - the red card they received as part of the setup. In reality they used a wet signature to represent this. This helped convey the concept that once a credential has been signed it is impossible to change the contents of that credential without invalidating the signature - providing integrity to the credential.

For simplicity the scenario did not represent individual attributes that credentials contained - for example a GMC credential might contain the doctors name and their GMC licence number. This was conveyed verbally instead.

During the session, the signature type - CL signatures, and the way that they work in the context of Verifiable Credentials was briefly touched upon. Due to the importance to understand why a doctor couldn't easily share a credential. Specifically the doctor contributes some secret information in a blind manner into the signing protocol as one of the credential attributes. Such that when the credential is signed and given to the doctor, only they can prove they know the secret value that was signed by the issuer. Even the issuer does not know this.

The wall of trusted DIDs and their corresponding public keys was used to discuss how the doctor, or rather their wallet, could verify the signatures on any credentials they were issued to ensure they were valid. The doctor is also capable of refusing a credential or suggesting changes before accepting - for example if the issuer had spelt their name wrong.

\subsubsection{Proof Request}

A proof request is an SSI interaction whereby an entity, generally referred to as a verifier, requests proof of certain identity information from a holder - in this scenario the doctor. The verifier is able to additionally specify the credential schema that the identity information should come from and the credential issuer if they wish.

For example, in the ecosystem modelled there were only two hospitals - lets call them Glasgow Hospital and Edinburgh hospital for simplicity, Edinburgh may request proof of a doctors name and DoB contained within an Identity Verification credential and only accept this proof if it was issued by Glasgow hospital (represented by it's DID).

This was a complex interaction that was challenging to represent within the scenario, so verbal communication was used to outline the majority of it. The actor representing the entity asking for a proof from the doctor, wrote this request on a piece of card which was then passed to the doctor. It was made clear that this communication went across an already established DID connection and that proof requests could include a subset of attributes from across multiple credentials.

\subsubsection{Credential presentation}

A credential presentation was the final SSI interaction used repeatedly throughout the scenario. This is the process by which a doctor, through the use of their digital wallet, responds to a proof request by creating a cryptographic presentation from one of more verifiable credentials within their wallet. 

Within the scenario, this was illustrated by filling out a card and creating a wet signature on this card. The card was then passed to the entity/actor that requested the proof request. Again, a lot of the complexity was conveyed verbally. It was explained that a credential presentation is not the same as giving the credential within a digital wallet to the entity requesting it. A presentation is a new and distinct cryptographic object, that is created from one or more credentials, and can contain any number of attributes from these credentials. This has both privacy and security benefits.

Going back to the hospital example, Edinburgh may request a doctor to prove their name, date of birth and GMC licence number when initially employing a new doctor. This proof request can be responded to in a single credential presentation that combines the attributes from two separate credentials originally issued by different entities into a new object that is still cryptographically verifiable using the public keys of the issuers. The keys that are publicly available and were stuck on the wall during the initial setup.

\subsection{Evaluating SSI design principles}

The last session of the workshop, involved presenting SSI in the context of traditional identity management systems such as federated and user centric identity management systems. 

Then asking for positive and negative aspects of the SSI system presented to them. Challenges to it's implementation were also collected through this process.

After this, eight design principles and their definitions selected from the literature were presented to the group. Mentimeter, a tool for audience engagement, was used to gauge how the participants valued these design principles. A couple of additional questions were also asked.

Before we introduced the design principles from the literature to our audience, we asked them what they thought was the most important feature of future technology that would help them trust it. The list included the following:

\begin{itemize}
    \item Use all over the ecosystem, all entities need to participate.
    \item Attention for end users, usability, convenience, workable.
    \item Buy-in from government and NHS.
    \item Future proof.
    \item Resilient, reliable, fraud resilient, protection, security.
    \item Control.
    \item Transparent data sharing, clear, clarity.
\end{itemize}{}

Comparing this list with the list of design principles that we distilled from the literature demonstrates that our audience adds two specific principles to the generic list. The first is that they find it important to know that all entities will be involved in the SSI ecosystem, including the government and the NHS. The success of the system depends on buy-in of all of the involved entities, and that is something that should be developed from the start. Second is the attention for usability and convenience. User engagement is important from the start and throughout the development process.  

\begin{figure}
    \centering
    \includegraphics[width=1.0\linewidth]{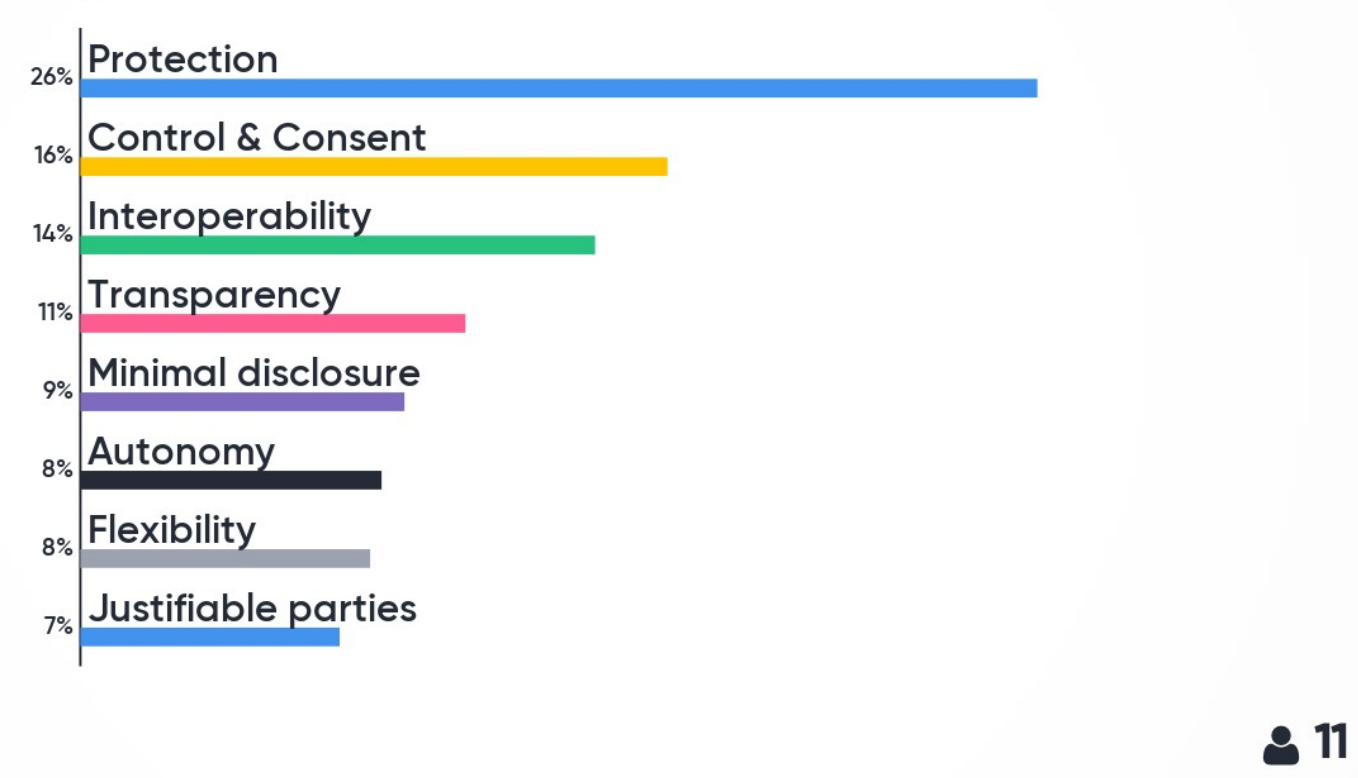}
    \caption{Principles ranked by importance}
    \label{fig:order_importance}
\end{figure}

\begin{figure}
    \centering
    \includegraphics[width=1.0\linewidth]{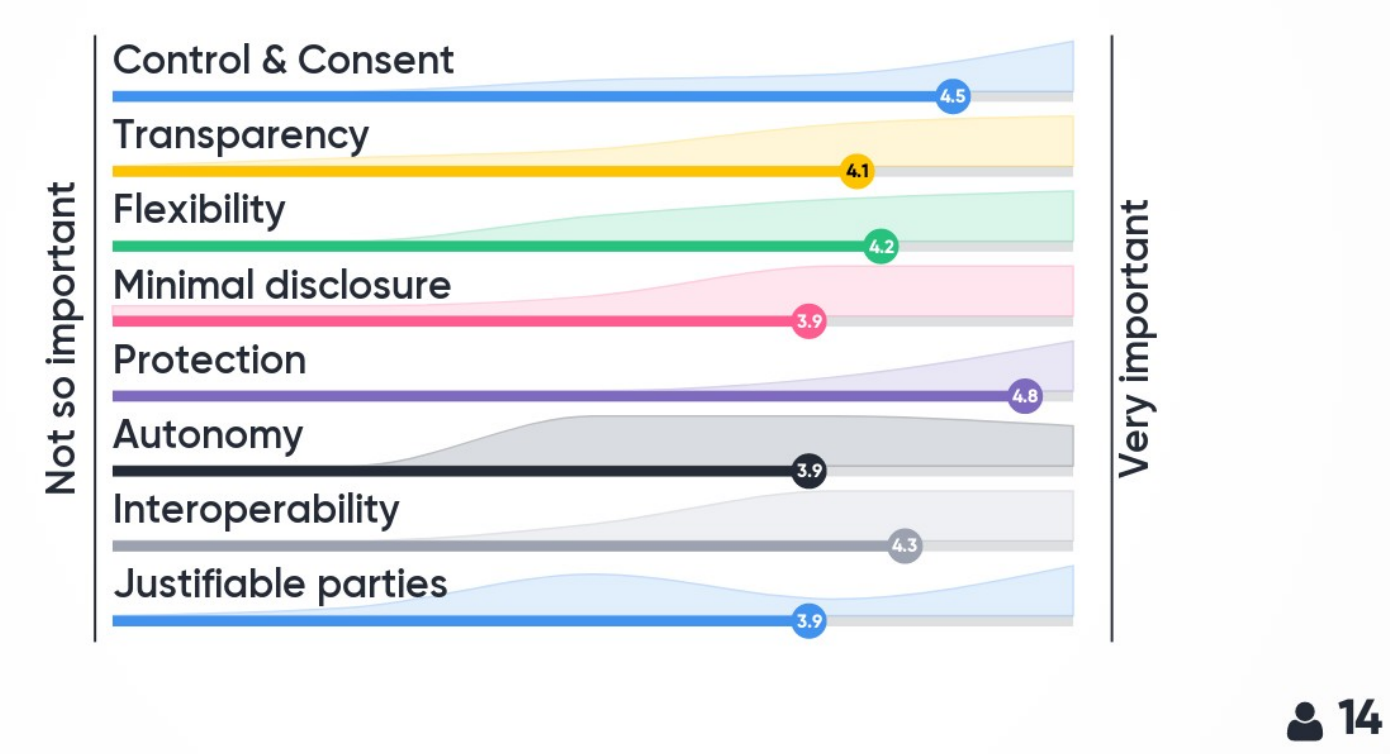}
    \caption{Principles individually rated}
    \label{fig:principle_importance}
\end{figure}

We selected eight principles from the literature and explained the definitions to the participants. Then we asked them to rank the principles in order of importance. The majority selected protection as the most important, followed by control \& consent and interoperability. Then we asked to rank importance for each individual principle. Again, the highest scores were for: protection; control and consent; and interoperability. Figures \ref{fig:order_importance} and \ref{fig:principle_importance} show the results of the Mentimeter polls. This was used to get a sense of the room and the figures should be interpreted with that in mind

\section{Workshop results}

The workshop led to a number of key learning elements about identity management for healthcare professionals. This can be generally summarised as being complex, fragmented; and time consuming in its current form. The process mapping session led to an increased understanding of the identity landscape within healthcare. We began the session with eight identity moments that we framed as being chronological, the idea behind this was to provide a rough skeleton for participants to expand on in their process maps. It came to light that we missed a key identity moment within a doctors career, namely appraisal and re-validation. A process whereby healthcare professionals must prove to relevant authorities that they have gone through the relevant training to keep their skills up to date such that they can still practice. This is a repetitive process that occurs every three years. It was interesting to find out that even the top level professionals in attendance typically spent a couple of days every three years getting their documents in order for this procedure. 

Another point of clarification was that while we positioned the 8 identity moments as chronological, a lot of them happen in parallel. For example, a medical student typically gets identity checked by the GMC prior to graduating, they also spend their final year applying for jobs so that on graduation they are ready to begin their career immediately. It was pointed out that there is no strict temporal relationship between the eight stages initially identified and a large part of the frustrations come from the fact that the majority of these stages are repeated over the course of a doctors career. Each time requiring the same time consuming procedure.

A good example of this comes from what we broadly termed doctor rotation, something commonly experienced within the healthcare system. Especially for young doctors who can rotate to a new location and role as often as every four months. While this is reduces significantly as doctors progress in their career, it still a common occurrence. Every time a doctor moves to a new hospital there are a number of tasks that the doctor needs to complete in order to on-board into the institution:
\begin{itemize}
    \item They must complete a full identity verification check to the NHS standards.
    \item They must provide evidence supporting all the claims they made in their application. This is typically verified by a consultant and their were estimates it takes up to a full day of their time.
    \item They must complete an induction session, taking between one and two days. This induction is required for both new locations (e.g. hospitals) and new trusts and generally includes repeated content due to lack of standardisation across locations and trusts.
    \item They must organise an appointment with occupational health, to prove they have up to date vaccinations. If they don't or are unable to prove they do then doctors must have their vaccinations refreshed. Often leading to needless re-vaccinations.
    \item 
\end{itemize}{}

Another big bugbear of the group, particularly from attendees still going through training, was keeping track of all the training events they had attended. Including the need to enter this information into multiple distinct silos. This was further exacerbated by frustrating user experiences, different document format requirements and even reviewers specifically asking for physical copies due to the added burden that reviewing digitally uploaded documents entailed. A clear example of how digital tools have failed healthcare professionals by increasing rather than reducing the burden placed on them to manage their professional careers.

To summarise, the process mapping and ensuing discussions highlighted numerous frustrations experienced by healthcare professionals just to meet the requirements for managing their career. A phrase that came up was the need to professionalise the digital experience throughout a doctors career.

The method for conveying the capabilities of SSI and how these might change the identity moments a doctor experiences throughout their career was a success. The majority of participants were engaged and achieved a good level of understanding as shown through the questions that this generated. Many of the participants indicated that they would be receptive to these changes being implemented, in particular a trainee at the RCPE was very supportive.

The workshops additionally surfaced a number of challenges that attendees thought would need be to overcome in order to roll this out within a healthcare system. Many participants were senior professionals within healthcare, so had experiences of other attempts to digitise aspects of healthcare. The three core challenges identified were:

\begin{itemize}
    \item \textbf{Funding and business model}: Who is paying for these tools and what is in it for them. There is no clear path to monetisation of the system, however for this to work it needs to be well funded in order to produce something that can scale. It was suggested that part of doctors annual fees and membership to organisations like the GMC and the Royal Colleges could be allocated towards a system like this.
    \item \textbf{Adoption}: An SSI system works only when it achieves large scale adoption. In order for this to happen it needs to be led on a national level and show the benefit to the entire ecosystem. It was also pointed out that while benefit to doctors is relatively easy to show it needs to be able to show benefit to the individual trusted entities. They need to be the ones advocating for this system not the doctors. All stakeholders must be clear on why this shift is happening and what the benefit is too them.
    \item \textbf{Overreaching}: This technology is new and relatively untested at scale. An interesting point was made that it is important to take small steps to prove its value and build human trust before expanding the scope. Attempting anything too large too quickly and failing could be disastrous for the trust placed in the system and underlying technology.
\end{itemize}{}


\section{Conclusion}

To conclude, participants were largely in favour of the technology described in the workshop. At least from the viewpoint of it's worth exploring further. A number of them were keen to be involved in further iterations, offering support finding additional doctors and medical students to further explore the requirements of any technical solution from their perspective. This research as further reinforced the importance of developing user driven systems, ensuring that any solution that does get rolled out is meeting the actual needs of those it is designed for. The research identified overlooked design principles, but also showed that the design principles commonly referred to within the academic literature, when explained, were also considered important by the system users within a healthcare context.

Next steps include developing a proof of concept which can then be validated with real users, this will include validating it against the properties deemed important by participants: Usability and Security. Additionally, A medical students identity interactions in themselves seem complex, it would be valuable to run another workshop specifically focusing on this area. This should provide a different perspective from that gained throughout this workshop.

The professionalisation of the digital experience within a doctors career is long overdue. The COVID-19 crisis has highlighted just how important these solutions can be for the any health service. Staffing needs fluctuate widely across the country, as different regions and hospitals hit their peaks of this crisis at different times. The ability to redeploy doctors to highly stressed areas within the service in minutes rather that days - See figure \ref{fig:id_moments}. While a portion of these induction processes can perhaps be ignored, trusts still have to meet strict legal requirements around identity verification \cite{ruscoe_identity_2018} - this all takes time which could be spent saving lives. In addition to the staffing flexibility an SSI based solution could bring, it also has huge potential to cut costs and increase efficiency. A report from the McKinsey Global Institute estimate that institutions requiring high assurance identification for on-boarding could see a 90 percent reduction in costs \cite{mckinsey_costs}. The World Economic forum estimates savings of up to \$205 Billion from seamless and secure sharing of medical information between healthcare organisations \cite{wef_costs}.

This seems to be the opportune moment to develop and deploy an SSI solution. However, it is imperative that any solution that is developed, especially if rushed through during a crisis, is clear about what it wants to achieve for who. Furthermore, these solutions should identify a process by which to validate how well they are meeting the predefined aims and requirements of those who are actually going to be using it - healthcare professionals.


\section*{Acknowledgements} {
The authors would like to extend their thanks to Dr Manreet Nijjar, an infectious disease consultant and co-founder of truu.id, a company working to realise much of what is discussed throughout this paper. His attendance at the workshop was invaluable. Also thanks to Professor Derek Bell and Pernille Marqvardsen without whom this workshop would not have been possible.
}

\bibliography{references}
\bibliographystyle{vancouver}

\end{document}